\documentclass[showpacs,preprint,preprintnumbers,amsmath,amssymb]{revtex4}
 \usepackage{dcolumn}
 \usepackage{bm}

\begin{document}

 \preprint{}

 \title{Topology of Knotted Optical Vortices }

\author{Ji-Rong Ren }
 \author {Tao Zhu }
\thanks{Corresponding author. Email : zhut05@lzu.cn }
 \author {Yi-Shi Duan}

\affiliation{Institute of Theoretical Physics, Lanzhou University,
Lanzhou 730000, P. R. China}

 \date{\today}

 \begin{abstract}
Optical vortices as topological objects exist ubiquitously in
nature. In this paper, by making use of the $\phi$-mapping
topological current theory, we investigate the topology in the
closed and knotted optical vortices. The topological inner structure
of the optical vortices are obtained, and the linking of the knotted
optical vortices is also given.
\end{abstract}



 \maketitle

 \section{Introduction}
Optical vortices\cite{opt1,opt2,opt3,optic}, which exist ubiquitously in
nature, are one of the phase singularities existed in light beams ,and are the
vortices of electromagnetic energy flow. It have drawn great interest and have
been studied intensively in many ways because it is of importance for
understanding fundamental physics and have many important applications. Optical
vortices are topological objects on wave-front surfaces and possess topological
charges which can be attributed to the helicoidal spatial structure of the
wave-front around a phase singularity. They occur when three or more complex
scalar light waves interfere. The total wave amplitude of interference field is
zero and the phase is undefined. Optical vortices are generically points in two
dimensional fields and form vortex lines in three dimensions. As topological
objects, the topological and geometrical properties of optical vortices have
been studied by many
physicists\cite{opttopo1,opttopo2,optknot1,optknot3,optknot4,optknot2}.

It has long been known that optical vortices can be closed. In 2001,
Berry and Dennis showed theoretically that specific superpositions
of beams could be generated, in which optical vortices could be
linked together and even knotted\cite{optknot1}. Since the original
analysis of knotted and linked optical vortex loops have been done,
there are a great deal of works on the knotted topology of the
optical vortices have been done\cite{optknot3,optknot4}. In the
laboratory, the stable link and knot structures have been
produced\cite{optknot2}. The observations of knotted optical
vortices demonstrate the precision control of light beams that is
required to create complex regions of zero light intensity. Such
regions could be used, for example,to confine cold atoms and
Bose-Einstein condensates in complex topologies. Knotted objects
have been studied theoretically since Lord Kelvin¡¯s vortex atom
hypothesis\cite{vortexatom} in a range of different physical
situations, such as hydrodynamics\cite{fluid}, field
theory\cite{feild,field2} and nonlinear excitable
media\cite{winfree}. The knotted optical vortex is a new type of
knotted object, the two or more such knots are called a link, i.e.,
a family of knots. It is known that for a knot family there are
important characteristic numbers to describe its topology, such as
the self-linking and linking numbers. So, it is very necessary to
use topological viewpoint to study these knot characteristics.

The $\phi$-mapping topological current theory\cite{topoduan} and the
decomposed theory of gauge potential\cite{quanvortex} play an
important role in studying the topological invariant and structure
of a physical system and have been used to study the topological
current of a magnetic monopole\cite{monopole}, the topological
string theory\cite{string}, the topological structure of the defects
of space-time in the early universe as well as its topological
bifurcation\cite{space}, the topological structure of the
Gauss-Bonnet-Chern theorem\cite{gaussmap}, the topological structure
of the London equation in a superconductor\cite{landon} and point
defect of a vector parameter\cite{point}.

In this paper, by making use of the so-called $\phi$-mapping
topological current theory, we study the topological current of the
optical vortices in three dimensional fields, and the topological
inner structure of these optical vortices are obtained. For the case
that the optical vortices are closed and knotted curves, we
discussed the topological invariant of these knotted family in
details. This paper is arranged as follows. In Sec.II, we study the
topological current density of the optical vortices, this
topological current density don't vanish only when the optical
vortices exist, the topological charge of optical vortices are
expressed by the topological quantum numbers,the Hopf indices and
Brouwer degrees of the $\phi$-mapping. In Sec.III, we introduce a
important topological invariant to describe the optical vortices
when they are linked and knotted, it is just the sum of all the
self-linking and all the linking numbers of the knot family. In Sec.
IV is our concluding remarks.

 \section{Topological charge current density of optical vortices}
It is known that the three-dimensional fields of interfering light
beams can be denoted by the complex scalar function
$\psi(\vec{r})=\psi(x^1,x^2,x^3)$, $\psi(\vec{r})$ is maps from
space to the complex numbers, so $\psi : R^3\rightarrow \mathcal
{C}$. The $\psi(\vec{r})$ is superpositions of free space optical
modes with the same frequency, so the optical vortices placed where
the phase singularities exist are temporally stable. In the
theoretically analysis done by Berry and Dennis\cite{optknot1},
$\psi(\vec{r})$ is the complex scalar solution of the Helmholtz
equation
\begin{equation}
\nabla^2\psi(\vec{r})+\psi(\vec{r})=0,
\end{equation}
and it possess optical vortices (phase singularities) in the form of
knots or links. In the laboratory\cite{optknot2}, the optical
vortices can be produced by the special superpositions of laser
beams (Laguerre-Gaussian beams), so the $\psi(\vec{r})$ is just this
superposition field. Here in our discussions, $\psi(\vec{r})$ is a
generic complex scalar optical field which possesses closed and
knotted optical vortices.

The complex scalar optical field $\psi(\vec{r})$ is space dependent,
and it can be written as
\begin{equation}
\psi(\vec{r})=\|\psi\|~
e^{i\chi}=\phi^1(\vec{r})+i\phi^2(\vec{r}),\label{wavefunction}
\end{equation}
where $\phi^1(\vec{r})$ and $\phi^2(\vec{r})$ can be regarded as
complex representation of a two-dimensional vector field
$\vec{\phi}=(\phi^1,\phi^2)$, $\|\psi\|=\sqrt{\psi^*\psi}$ is the
modulus of $\psi$, and $\chi$ is the phase factor. The current
density $\vec{J}$ associated with $\psi(\vec{r})$ is defined as
\begin{equation}
\vec{J}=Im (\psi^* \nabla \psi)=\|\psi\|^2 \vec{v},\label{current}
\end{equation}
the $\vec{v}$ is the velocity field. From the expressions in
Eq.(\ref{wavefunction}) and Eq.(\ref{current}), the velocity field
$\vec{v}$ can be rewritten as
\begin{equation}
\vec{v}=\frac{1}{2i}\frac{1}{\|\psi\|^2}(\psi^*\nabla\psi-\nabla\psi^*\psi)=\nabla\chi,\label{velocity}
\end{equation}
it becomes the gradient of the phase factor $\chi$. The vorticity
field $\vec{\Omega}$ of the velocity field $\vec{v}$ is defined as
\begin{equation}
\vec{\Omega}=\frac{1}{2\pi}\nabla\times\vec{v}.\label{vorticity}
\end{equation}
Form Eq.(\ref{velocity}),we directly obtain a trivial curl-free
result: $\vec{\Omega}=\frac{1}{2\pi}\nabla\times\nabla\chi=0$. But
in topology, because of the existence of  optical vortices in the
interference optical fields $\psi$, the vorticity $\vec{\Omega}$
does not vanish\cite{quanvortex}. So in the following discussions,
we will study that what the exact expression for $\vec{\Omega}$ is
in topology.

Introducing the unit vector $n^a=\phi^a/\|\phi\| (a=1, 2; n^an^a=1)$,one can
reexpress the velocity field $\vec{v}$ as
\begin{equation}
\vec{v}=\epsilon_{ab}n^a\nabla n^b,
\end{equation}
and the vorticity $\vec{\Omega}$ is
\begin{equation}
\Omega^i=\frac{1}{2\pi}\epsilon^{ijk}\epsilon_{ab}\partial_jn^a\partial_kn^b,~~~~~i,j,k=
1, 2, 3.\label{topovorticity}
\end{equation}
Obviously, it is just the topological current of the optical
vortices in three dimensional optical fields.

Using the $\phi$-mapping theory\cite{topoduan}, the topological
current $\vec{\Omega}$ is rewritten as
\begin{equation}
\Omega^i=\delta^2(\vec{\phi})D^i(\frac{\phi}{x}),\label{delat}
\end{equation}
where the $D^i(\frac{\phi}{x})$ is the vector Jacobians of
$\psi(\vec{r})$, and it is defined as
\begin{equation}
D^i(\frac{\phi}{x})=\frac{1}{2}\epsilon^{ijk}\epsilon_{ab}\partial_j\phi^a\partial_k\phi^b,
\end{equation}
we can see from the expression (\ref{delat}) that the vorticity
$\vec{\Omega}$ is non-vanishing only if $\vec{\phi}=0$, i.e., the
existence of the optical vortices, so it is necessary to study these
zero solutions of $\vec{\phi}$. In three dimensions space, these
solutions are some isolate zero lines, which are the so-called
optical vortices in three dimensional space.

Under the regular condition $$D^i(\phi/x)\neq0,$$ the general
solutions of
\begin{equation}
\phi^1(x^1,x^2,x^3)=0,~~~~~ \phi^2(x^1,x^2,x^3)=0
\end{equation}
can be expressed as
\begin{equation}
x^1=x^1_k(s), ~~~x^2=x^2_k(s),~~~x^3=x^3_k(s),
\end{equation}
which represent $N$ isolated singular strings $L_k$ with string
parameter $s$ $(k=1,2,\cdots,N)$. These singular strings solutions
are just the optical vortices solutions in three dimensions space.

In $\delta$-function theory\cite{deltfun}, one can obtain in three
dimensions space
\begin{equation}
\delta^2(\vec{\phi})=\sum_{k=1}^N \beta_k
\int_{L_k}\frac{\delta^3(\vec{x}-\vec{x}_k()s)}{|D(\frac{\phi}{u})|_{\Sigma_k}}ds,\label{delta3}
\end{equation}
where
$$D(\frac{\phi}{u})|_{\Sigma_k}=\frac{1}{2} \epsilon^{jk}\epsilon_{mn}\frac{\partial\phi^m}{\partial
u^j}\frac{\partial\phi^n}{\partial u^k},$$ and $\Sigma_k$ is the
$k$th planar element transverse to $L_k$ with local coordinates
$(u^1,u^2)$. The $\beta_k$ is the Hopf index of $\phi$ mapping,
which means that when $\vec{x}$ covers the neighborhood of the zero
point $\vec{x}_k(s)$ once, the vector field $\phi$ covers the
corresponding region in $\phi$ space $\beta_k$ times. Meanwhile the
direction vector of $L_k$ is given by
\begin{equation}
\frac{dx^i}{ds}|_{x_k}=\frac{D^i(\phi/x)}{D(\phi/u)}|_{x_k}.\label{direction}
\end{equation}
Then from Eq.(\ref{delta3}) and Eq.(\ref{direction}) one can obtain
the inner structure of $\Omega^i$:
\begin{equation}
\Omega^i=\sum_{k=1}^N W_k
\int_{L_k}\frac{dx^i}{ds}\delta^3(\vec{x}-\vec{x}_k(s))ds,\label{topo}
\end{equation}
where $W_k=\beta_k\eta_k$ is the winding number of $\vec{\phi}$
around $L_k$, with $\eta_k=sgn D(\phi/u)|_{\vec{x}_k}=\pm1$ being
the Brouwer degree of $\phi$ mapping. The sign of Brouwer degrees
are very important, the $\eta_k=+1$ corresponds to the vortex, and
$\eta_k=-1$ corresponds to the antivortex. The integer number $W_k$
measures windings of the phase around the phase singularities, and
is called the topological charge  of the optical vortex.  Hence the
topological charge of the optical vortices $L_k$ is\cite{feild}
\begin{equation}
Q_k=\int_{\Sigma_k}\Omega^id\sigma_i=W_k.\label{topochar}
\end{equation}
In the theory of the optical vortices, the topological charge plays
the role of an angular momentum. The Eq.(\ref{topochar}) shows us
that the topological current $\vec{\Omega}$ describes the density of
optical vortices in space. So, we call the topological current
$\vec{\Omega}$ the topological charge current density  of the
optical vortices.

The results in this section show us the topological inner structure
of the topological charge current density $\vec{\Omega}$. The
topological charge of the $k$th optical vortex in three dimensions
can be expressed by the topological numbers:
$Q_k=W_k=\beta_k\eta_k$, and the $\eta_k=+1$ corresponds to the
vortex, and $\eta_k=-1$ corresponds to the antivortex.

 \section{Linking numbers of knot family}

Topology has played a very important role in understanding the knot
configurations, so it is necessary to study the topology in the knotted optical
vortices. In order to do that, we define a the helicity integral\cite{heli}
\begin{equation}
H=\frac{1}{4\pi^2}\int \vec{v}\cdot\nabla\times\vec{v}d^3x,
\end{equation}
this is an important topological knot invariant and it measures the
linking of the optical vortices. From the Eq.(\ref{vorticity}), the
helicity integral can be changed as
\begin{equation}
H=\frac{1}{2\pi}\int \vec{v}\cdot\vec{\omega}d^3x.\label{helint}
\end{equation}
For the magnetic helicity\cite{magnheli}, $\vec{v}$ is the magnetic
field $\vec{B}$ and $\vec{\Omega}$ the $\vec{A}$, the helicity
integral measures the linking number of the field lines, averaged
over all pairs of field lines, and weighted by magnetic flux. Anther
application of the helicity is in fluid mechanics\cite{fluid},
helicity $H$ measures the linking of the fluid vortex lines. In this
paper, we apply this helicity to study the linking of the optical
vortices. In Section.II, we have known that the vorticity
$\vec{\Omega}$ does not vanish only when the optical vortices exist,
so from the Eq.(\ref{helint}), when the optical vortices exist, the
helicity also does not vanish. So helicity is just the optical
vortices helicity.

Substituting Eq. (\ref{topo}) into Eq. (\ref{helint}), one can
obtain
\begin{equation}
H=\frac{1}{2\pi}\sum_{k=1}^{N}W_k \int_{L_k}\vec{v}\cdot
d\vec{x}\label{helint1},
\end{equation}
for closed and knotted lines, i.e., a family of knots $\xi_k (k=1,
2, \ldots, N)$, Eq. (\ref{helint1}) becomes
\begin{equation}
H=\frac{1}{2\pi}\sum_{k=1}^{N}W_k \oint_{\xi_k}\vec{v}\cdot
d\vec{x}.
\end{equation}

It is well known that many important topological numbers are related
to a knot family such as the self-linking number and Gauss linking
number. In order to discuss these topological numbers of knotted
optical vortices, we define Gauss mapping:
\begin{equation}
\vec{m}: S^1 \times S^1 \rightarrow S^2,
\end{equation}
where $\vec{m}$ is a unit vector
\begin{equation}
\vec{m}(\vec{x},
\vec{y})=\frac{\vec{y}-\vec{x}}{|\vec{y}-\vec{x}|},\label{unit}
\end{equation}
where $\vec{x}$ and $\vec{y}$ are two points, respectively, on the
knots $\xi_k$ and $\xi_l$ (in particular, when $\vec{x}$ and
$\vec{y}$ are the same point on the same knot $\xi$ , $\vec{n}$ is
just the unit tangent vector $\vec{T}$ of $\xi$ at $\vec{x}$ ).
Therefore, when $\vec{x}$ and $\vec{y}$ , respectively, cover the
closed curves $\xi_k$ and $\xi_l$ once, $\vec{n}$ becomes the
section of sphere bundle $S^2$. So, on this $S^2$ we can define the
two-dimensional unit vector $\vec{e}=\vec{e}(\vec{x}, \vec{y})$.
$\vec{e}$, $\vec{m}$ are normal to each other, i.e. ,
\begin{eqnarray}
&&\vec{e}_1\cdot\vec{e}_2=\vec{e}_1\cdot\vec{m}=\vec{e}_2\cdot\vec{m}=0,
\nonumber\\&&\vec{e}_1\cdot\vec{e}_1=\vec{e}_2\cdot\vec{e}_2=\vec{m}\cdot\vec{m}=1.
\end{eqnarray}
In fact, the velocity field $\vec{v}$ can be decomposed in terms of
this two-dimensional unit vector $\vec{e}$:
$v_i=\epsilon_{ab}e^a\partial_i e^b-\partial_i\theta$, where
$\theta$ is a phase factor\cite{quanvortex}. Since one can see from
the expression $\vec{\Omega}=\frac{1}{2\pi}\nabla\times \vec{v}$
that the $(\partial_i\theta)$ term does not contribute to the
integral $H$, $v_i$ can in fact be expressed as
\begin{equation}
v_i=\epsilon_{ab}e^a\partial_ie^b.
\end{equation}
Substituting it into Eq.(14), one can obtain
\begin{equation}
H=\frac{1}{2\pi}\sum_{k=1}^{N}W_k
\oint_{\xi_k}\epsilon_{ab}e^a(\vec{x},
\vec{y})\partial_ie^b(\vec{x}, \vec{y})dx^i.\label{hel}
\end{equation}
Noticing the symmetry between the points $\vec{x}$ and $\vec{y}$ in
Eq.(\ref{unit}), Eq.(\ref{hel}) should be reexpressed as
\begin{equation}
H=\frac{1}{2\pi}\sum_{k, l=1}^N W_k W_l \oint_{\xi_k}
\oint_{\xi_l}\epsilon_{ab}\partial_i e^a\partial_j e^bdx^i\wedge
dy^j.\label{hel2}
\end{equation}
In this expression there are three cases: (1) $\xi_k$ and $\xi_l$
are two different optical vortices $(\xi_k\neq\xi_l)$, and $\vec{x}$
and $\vec{y}$ are therefore two different points
$(\vec{x}\neq\vec{y})$; (2) $\xi_k$ and $\xi_l$ are the same optical
vortices $(\xi_k=\xi_l)$, but $\vec{x}$ and $\vec{y}$ are two
different points $(\vec{x}\neq\vec{y})$; (3) $\xi_k$ and $\xi_l$ are
the same optical vortices $(\xi_k=\xi_l)$, and $\vec{x}$ and
$\vec{y}$ are the same points $(\vec{x}=\vec{y})$. Thus,
Eq.(\ref{hel2}) can be written as three terms:
\begin{eqnarray}
&&H=\sum_{k=1(k=l, \vec{x\neq}\vec{y})}^N \frac{1}{2\pi}W_k^2
\oint_{\xi_k} \oint_{\xi_k} \epsilon_{ab} \partial_i e^a\partial_j
e^b dx^i \wedge dy^j \nonumber\\&&+\frac{1}{2\pi}\sum_{k=1}^N W_k^2
\oint_{\xi_k} \epsilon_{ab} e^a\partial_i e^bdx^i
\nonumber\\&&+\sum_{k, l=1(k\neq l)}^N \frac{1}{2\pi}W_k W_l
\oint_{\xi_k} \oint_{\xi_l} \epsilon_{ab}
\partial_i e^a\partial_j e^b
dx^i \wedge dy^j.\label{hel3}
\end{eqnarray}
By making use of the relation
$\epsilon_{ab}\partial_ie^a\partial_je^b=\frac{1}{2}\vec{m}\cdot(\partial_i\vec{m}\times\partial_j\vec{m})$\cite{rela},
the Eq.(\ref{hel3}) is just
\begin{eqnarray}
H&=&\sum_{k=1(\vec{x}\neq\vec{y})}^N \frac{1}{4\pi}W_k^2
\oint_{\xi_k}\oint_{\xi_k} \vec{m}^*(dS) \nonumber \\ &&
+\frac{1}{2\pi}\sum_{k=1}^N W_k^2 \oint_{\xi_k} \epsilon_{ab}e^a
\partial_ie^bdx^i \nonumber \\ &&
+\sum_{k, l=1(k\neq l)}^N \frac{1}{4\pi} W_k W_l
\oint_{\xi_k}\oint_{\xi_l} \vec{m}^*(dS),\label{hel4}
\end{eqnarray}
where
$\vec{m}^*(dS)=\vec{m}\cdot(\partial_i\vec{m}\times\partial_j\vec{m})dx^i\wedge
dy^j(\vec{x}\neq\vec{y})$ denotes the pullback of the $S^2$ surface
element.

In the following we will investigate the three terms in the
Eq.(\ref{hel4}) in detail. Firstly, the first term of
Eq.(\ref{hel4}) is just related to the writhing number\cite{writnum}
$Wr(\xi_k)$ of $\xi_k$
\begin{equation}
Wr(\xi_k)=\frac{1}{4\pi}\oint_{\xi_k}\oint_{\xi_l}
\vec{m}^*(dS).\label{writhing}
\end{equation}
For the second term, one can prove that it is related to the
twisting number $Tw(\xi_k)$ of $\xi_k$
\begin{eqnarray}
\frac{1}{2\pi}\oint_{\xi_k}\epsilon_{ab}e^a\partial_ie^bdx^i
&&=\frac{1}{2\pi}\oint_{\xi_k}(\vec{T}\times\vec{V})\cdot
d\vec{V}\nonumber\\&&=Tw(\xi_k),\label{twisting}
\end{eqnarray}
where $\vec{T}$ is the unit tangent vector of knot $\xi_k$ at
$\vec{x}$ ($\vec{m}=\vec{T}$ when $\vec{x}=\vec{y}$) and $\vec{V}$
is defined as
$e^a=\epsilon^{ab}V^b(\vec{V}\perp\vec{T},\vec{e}=\vec{T}\times\vec{V})$.
In terms of the White formula\cite{writfor}
\begin{equation}
SL(\xi_k)=Wr(\xi_k)+Tw(\xi_k),\label{self}
\end{equation}
we see that the first and the second terms of Eq.(\ref{hel4}) just
compose the self-linking numbers of knots.

Secondly, for the third term, one can prove that
\begin{eqnarray}
&&\frac{1}{4\pi}\oint_{\xi_k}\oint_{\xi_l}
\vec{m}^*(dS)\nonumber\\&&=\frac{1}{4\pi}\epsilon^{ijk}\oint_{\xi_k}dx^i\oint_{\xi_l}dy^j
\frac{(x^k-y^k)}{\|\vec{x}-\vec{y}\|^3}\nonumber\\&&=Lk(\xi_k,\xi_l)~~(k\neq
l),\label{linking}
\end{eqnarray}
where $Lk(\xi_k,\xi_l)$ is the Gauss linking number between $\xi_k$
and $\xi_l$\cite{writnum}. Therefore, from Eqs.(\ref{writhing}),
(\ref{twisting}), (\ref{self}) and (\ref{linking}), we obtain the
important result:
\begin{equation}
H=\sum_{k=1}^N W_k^2 SL(\xi_k)+\sum_{k, l=1(k\neq
l)}^NW_kW_lLk(\xi_k, \xi_l).
\end{equation}
This precise expression just reveals the relationship between $H$
and the self-linking and the linking numbers of the optical vortices
knots family\cite{writnum}. Since the self-linking and the linking
numbers are both the invariant characteristic numbers of the optical
vorties knots family in topology, $H$ is an important topological
invariant required to describe the linked optical vortices in
optical systems.

 \section{Conclusion}
In this paper, we consider the topology of the closed and knotted
optical vortices in optical systems. By making use of the
$\phi$-mapping topological current theory, we obtain the topological
inner structure of the optical vortices, the topological charge of
the vortices can be expressed by the topological numbers: Hopf
indices and Brouwer degrees. Furthermore for the closed and knotted
optical vortices, we introduce the helicity to study the knotted
topology, the helicity is a topological invariant of the knots
family, and we find that it is just the total sum of all the
self-linking and all the linking number of the knotted optical
vortices family.

 \begin{acknowledgments}
This work was supported by the National Natural Science Foundation
of China.
 \end{acknowledgments}

 \end{document}